\pdfoutput=1
\documentclass[11pt]{article}

\usepackage{latex/acl}

\usepackage{times}
\usepackage{latexsym}

\usepackage[T1]{fontenc}
\usepackage[utf8]{inputenc}

\usepackage{microtype}

\usepackage{inconsolata}
\usepackage{graphicx}
\usepackage{graphics}
\usepackage{soul}
\usepackage{listings}

\usepackage{url}

\usepackage{breakurl}
\usepackage[breaklinks]{hyperref}

\title{Testing the Effect of Code Documentation on Large Language Model Code Understanding}
\author{William Macke \and Michael Doyle \\
  The MITRE Corporation\thanks{Approved for Public Release; Distribution Unlimited. Public Release Case Number 23-4132. Copyright \textcopyright 2024  The  MITRE  Corporation. ALL RIGHTS RESERVED.}
  \\
  \normalsize \texttt{\{wmacke,mdoyle\}@mitre.org}\\}
\definecolor{codeblue}{RGB}{51,166,156}
\definecolor{codered}{RGB}{211,54,130}
\definecolor{codekeywords}{RGB}{203,75,21}
\definecolor{backcolour}{RGB}{253,246,227}
\lstdefinestyle{mystyle}{
  backgroundcolor=\color{backcolour},
  commentstyle=\color{codeblue},
  keywordstyle=\color{codekeywords},
  stringstyle=\color{codered},
  basicstyle=\ttfamily\footnotesize,
  breakatwhitespace=false,         
  breaklines=true,                 
  captionpos=b,                    
  keepspaces=true,                 
  numbersep=5pt,                  
  showspaces=false,                
  showstringspaces=false,
  showtabs=false,                  
  tabsize=2,
  frame=single,
}

\begin{document}
\maketitle
\begin{abstract}
Large Language Models (LLMs) have demonstrated impressive abilities in recent years with regards to code generation and understanding. However, little work has investigated how documentation and other code properties affect an LLM's ability to understand and generate code or documentation. We present an empirical analysis of how underlying properties of code or documentation can affect an LLM's capabilities.  We show that providing an LLM with "incorrect" documentation can greatly hinder code understanding, while incomplete or missing documentation does not seem to significantly affect an LLM's ability to understand code.
\end{abstract}

% TODO: Add MITRE Copyright

\section{Introduction}
%@Chris
% \wm{A thought occurs, it may be a good idea to state our hypotheses up front here. Maybe something along the lines of 1. Providing documentation can improve an LLM's code understanding, 2 incorrect documentation hurts an LLM's code understanding and 3. an LLM is able to generate documentation that can effectively help its own code understanding (assuming these were hypothese we had going in to run the experiments. I definitely had 1 and 2 going into the experiments, @Pranay/Jacob/Chris did you all have hypothesis 3 going into these experiments?)}

% Did this ^ for ours with one sentence in the introduction - mdoyle

Recently, Large Language Models (LLMs) have approached or pushed the state of the art for multiple natural language processing (NLP) tasks and benchmarks such as machine translation (MT)~\cite{moslem-etal-2023-adaptive, Jiao2023-xp}, human evaluation of MT~\cite{kocmi-federmann-2023-large}, sentence completion, and question answering~\cite{OpenAI2023-xj}. The same can be said for some programming language processing (PLP) tasks, such as code generation~\cite{OpenAI2023-xj} and code translation~\cite{Pan2023-rx}, but other tasks, such as code summarization, have proven to be difficult for LLMs~\cite{Sun2023-sp}. These tasks requiring code understanding pose challenges distinct from those faced in natural language. As an example, a unique challenge in PLP is the rigidity of syntax and semantic precision required in generation or translation that is not required to the same degree in NLP. Solving these problems, or at least improving their solutions, has the potential to greatly increase productivity and satisfaction in software development, as has already been shown with the use of GitHub Copilot~\cite{githubResearchQuantifying}.

%One of the most important aspects in text-to-code generation is an accurate instruction of what the resultant software should do, \md{in the form of natural language descriptions or chain of thought reasoning}~\cite{structuredCoT}. 
While several works have shown the importance of applying effective prompting strategies for PLP with LLMs~\cite{le2023codechain, 10.1145/3540250.3549113, shinn2023reflexion}, very little work has investigated how the reliability of the documentation at input can affect an LLM's performance. In this work, we provide a preliminary empirical analysis of how documentation quality can influence this ability. We hypothesize that correct code documentation will improve an LLM's code understanding, and that its understanding will decrease as the prevalence and accuracy of the documentation is decreased. To the authors' knowledge this is the first work to consider code documentation reliability for this problem.

The rest of the paper is organized as follows: First we discuss other works and how they relate to our study. Then we present experimental analysis and results. Our experiments provide a basic analysis of how an LLM's code understanding performance degrades along with documentation quality and quantity, and we end with a discussion and analysis of the results.

\section{Related Works}
Code analysis is a very well-studied problem. Traditional compilers and syntax-tree parsers are commonly used to perform code analysis~\cite{4362893,lenarduzzi2020survey} or extract code metrics~\cite{1702388, timoteo2008software, agnihotri2020systematic}. However, these tools lack an LLM's ability to process natural language, and therefore are incapable of extracting semantic understanding or considering code documentation. Since LLMs have become widely available, many works have also attempted to leverage language models to analyze code. Several works have sought to create LLMs designed for code understanding tasks such as code generation, code completion, program repair, and code translation~\cite{xia2022practical, wang2023codet5+, bui2023codetf}. Unlike these works, instead of developing new tools we provide a rigorous analysis of where and how existing tools can best be leveraged. Our work also introduces other means of determining code understanding than the benchmarks developed in \citet{lu2021codexglue}.

Other works provided empirical analyses of existing models' abilities to analyze code. \citet{xu2022systematic} present an analysis of the performance of various LLMs on different code benchmarks, and introduce a new LLM. \citet{ma2023chatgpt} present an empirical analysis of using GPT to generate syntax trees, call graph and other syntactic and semantic representations of code. They conclude that GPT has approximately the same abilities as an abstract syntax tree (AST) parser. Similarly \citet{palacio2023evaluating} introduce a framework called ASTxplainer that uses questions about an AST as an evaluation criteria for LLMs. \citet{leinonen2023comparing} provide an empirical study comparing human generated explanations with those generated by GPT. All of these works are focused on evaluating an LLMs performance in general and, unlike this work, none investigate how code content or documentation can effect an LLM's understanding.

% TODO: double check above

% TODO: maybe add a section on LLM unit test generation?

\section{LLM Code Understanding}
%TODO: rewrite to improve flow

It has been argued that \textit{understanding} is merely the ``knowledge of causes''~\cite{Pritchard2014}, but one can see that this is an incomplete definition of what it means to truly understand something and that ``a proper explanatory grip on how cause and effect are related''~\cite{Pritchard2016-PRISIF} is required for true understanding. Nevertheless, the knowledge of causes is at least a crucial early step in developing a complete understanding of a topic.

If this is the basic definition we are using for our basis of determining understanding, then to truly assess whether a piece of code is \textit{understood} by a person or a language model, then the subject must display knowledge of the cause and effect relationship of the code. Or, to put more simply, the subject should be able to answer the question, ``What does this piece of code do?''

Past work has attempted to test an LLM's ability to answer this question through code translation by measuring errors in the target language~\cite{Pan2023-rx}; through code summarization by measuring differences in BLEU, METEOR, or ROUGE-L scores between an LLM-generated summarization and the reference summarization~\cite{Sun2023-sp}; and through the generation of unit tests by measuring statement and branch coverage across various JavaScript libraries~\cite{schafer2023empirical}. Our approach for testing an LLM's understanding of source code is most closely related to the latter-most method, but differs by not improving upon the initial attempt at unit test generation iteratively if the tests fail.

%\subsection{Understanding through Verification}
% Talk about formal verification?
%TODO: cite sketch paper on program specification as input output pairs
% ^ Looks like you did that in paragraph 2 - mdoyle
% This section is very good and succinct. I just added a little bit to tie it back to documentation/comments.

% One of the core requirements for code understanding is predicting a program's behavior. This can range from giving loosely defined natural language descriptions of what a piece of software does to writing a formal specification~\cite{}.

% TODO: ^ Cite

In many cases, a program can be represented as a mapping from inputs into outputs~\cite{solar2009sketching}. For example, the program $square(n)$ can be described with $square(1)=1$, $square(2)=4$,... etc. Unsurprisingly then, how well an individual can predict input-output pairs can serve as a very useful surrogate metric for how well an individual understands a piece of code. After all, someone cannot truly understand software without being able to predict the software's behavior. Module documentation, function docstrings, inline code comments, and variable names provide hints or detailed descriptions of what the user can expect of these input-output pairs, and often allow human users to better predict that behavior.

Modern unit test frameworks provide both a formal language for describing input-output pairs and an easily automated method of checking an LLM's output. In order to measure an LLM's ability to generate these input-output pairs, we therefore task the LLM with generating unit tests for a piece of software while varying the quality or quantity of the documentation at input, as well as other code properties. We then measure the percent of unit tests that pass vs the number that fail or produce an error as the percent of correctly predicted input-output pairs.  

\section{Experimental Setup}
In the following section we describe our experimental process for testing an LLM's ability to understand code well enough to produce successful unit tests. In particular, we present an experimental setup for testing the generation of unit tests under varying initial documentation conditions.

%\subsection{Testing Understanding through Verification}
%As mentioned above, we use unit tests as a framework for generating/automatically testing input-output pairs predicted by an LLM. 
To test an LLM's ability to understand a piece of software, we task the language model to generate unit tests. We then run the unit tests and record each test as 1 of 3 results: \textbf{Runtime Error}, where a unit test crashed before it could finish running; \textbf{Failure}, where a unit test ran but failed to pass; and \textbf{Success}, where a unit test ran and successfully passed. All LLMs used their default parameters during experiments.
%In addition, we also measure both line and code coverage of the generated unit tests to ensure trivial unit tests aren't being generated (e.g. \emph{assert true}).
%Will add the above if we have space/time

We use all 164 HumanEval~\cite{chen2021codex} ground truth solutions as our basis for generating unit tests.~\footnote{Available for use under the MIT license \href{https://github.com/openai/human-eval}{here}.} We chose HumanEval both for its common use as an LLM benchmark for code understanding~\cite{hong2023metagpt,zelikman2023parsel,wang2023intervenor} (although intended for text-to-code generation), and for its wide variety of self encapsulated functions that can be easily run within unit test frameworks. We perform several post processing steps on the LLM generated code to ensure it can run automatically. First, we remove all import statements that result in an error (the LLM will commonly import hallucinated modules as it was not provided the original module names at input). Second, we append the LLM's code to a file with the HumanEval function to ensure the unit test has access to all code pieces required to run.

We generate several automatic variations of each function in the HumanEval dataset. \textbf{Base File}, which includes only the ground truth solution for a HumanEval function with no docstring comments. \textbf{Comments}, which is the \textit{base file} but includes any comments and the docstring provided for the HumanEval function. The docstrings for HumanEval functions typically contain a description of the function along with a few execution examples. \textbf{Random Comments} which has the \textit{base file} but with a random comment or docstring from a different HumanEval function, \textbf{Animal Variable Names}, which is the \textit{base file} but with all variable and (non-external) function names replaced with animals  (e.g. Bird, Cat, Dog etc.), and \textbf{Random Variable Names} which is the \textit{base file} but with all variable and function names replaced with random strings. For instance, a variable \lstinline{format_str} might be changed to \lstinline{eMbafsd}. In addition, we also investigate how removing random portions of the docstrings, line-by-line, can effect performance. Each line was kept with uniform random probability. We evaluate unit tests generated from 10\%, 25\%, 50\% and 75\% of the original docstrings remaining. To avoid any infinite loops generated by the LLM, all tests would automatically fail after taking more than 10 seconds to run. Figure~\ref{fig:example_humaneval} shows an example of a function from the \textbf{Comments} category.

In order to determine that the LLMs are not generating trivial tests, we also conduct an analysis of how much of the source code is executed by the generated unit tests. We use line coverage, or the percent of lines in the source code that are executed by the unit tests, as our metric for this analysis.

% TODO: Put a quote of the prompt used somewhere (either experiment or results section)
\begin{figure}
    \centering
\begin{lstlisting}[language=Python, showstringspaces=false]
def monotonic(l: list):
    """Return True is list elements 
    are monotonically increasing
    or decreasing.
    >>> monotonic([1, 2, 4, 20])
    True
    >>> monotonic([1, 20, 4, 10])
    False
    >>> monotonic([4, 1, 0, -10])
    True
    """
    if l == sorted(l) or\ 
        l == sorted(l, reverse=True):
        return True
    return False    
\end{lstlisting}
\caption{Example HumanEval reference implementation with docstring.}
\label{fig:example_humaneval}
\end{figure}

\section{Results}

% \wm{In this section we present results from the experimental setup described above.}

% \subsection{Verification Results}
% commenting out because only one section now - mdoyle

\begin{figure}
    \centering
    \includegraphics[width=\linewidth]{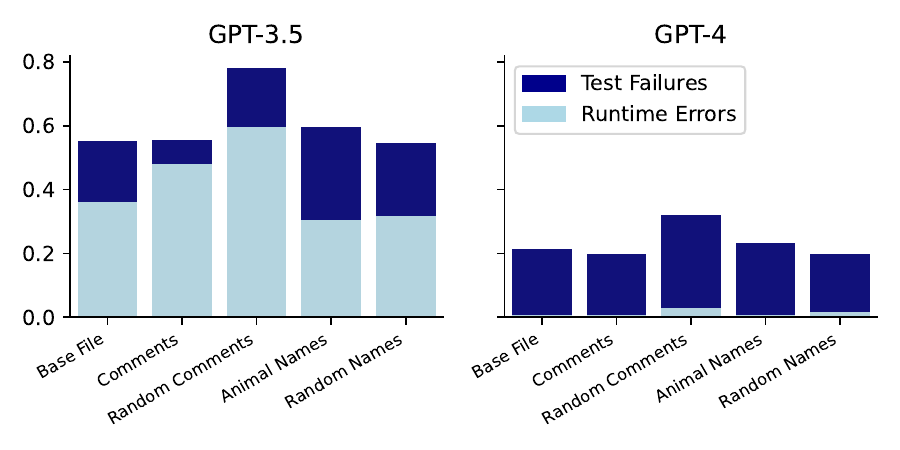}
    \caption{Proportion of runtime errors or failed tests that happen with GPT-3.5 (left) and GPT-4 (right) generating unit tests on modified versions of HumanEval code.}
    \label{fig:humaneval_cat}
\end{figure}

Figure~\ref{fig:humaneval_cat} shows the results of GPT-3.5 (\textit{gpt-3.5-turbo}) and GPT-4 generating unit tests on the variations of the HumanEval dataset discussed above.~\footnote{These results were generated using OpenAI's~\href{https://platform.openai.com/docs/guides/text-generation/chat-completions-api}{Chat Completions API}. OpenAI's sharing and publication policy regarding the use of their API can be seen \href{https://openai.com/policies/sharing-publication-policy\#research}{here}.} There are several key things to note from these results. First the number of runtime errors produced by GPT-3.5 is much greater than those produced by GPT-4, as is consistent with many results for similar code understanding tasks~\cite{OpenAI2023-xj}. A manual examination of the generated test cases shows that GPT-3.5 often generated simple assert statements rather than using pytest, and many of the runtime errors were caused by assert statements failing. Second, we note that the \textit{random comments} scenario performs worse than any other scenario on both GPT-3.5 and 4 (only 22.1\% and 68.1\% successes respectively) as shown in Figure~\ref{fig:humaneval_cat}, confirming our hypothesis that incorrect documentation can hurt an LLM's understanding of a piece of code. A statistical bootstrap shows that \textit{random comments} have a higher proportion of runtime errors and failed tests than any other version of the code by a statistically significant margin with $\alpha=0.05$. Thirdly, we note that changing the code content (via variable names), had a relatively minor effect on the LLM's code understanding when compared with the \textit{base file}. A bootstrap shows that changing variables to \textit{random names} did not impact the proportion of errors and failed tests by a statistically significant margin, but changing the variables to \textit{animals} did cause a statistically significant change (albeit a small one of 44.7\% to 40.6\% successes for GPT-3.5 and 78.5\% to 76.6\% for GPT-4). Finally, having comments did not \textbf{significantly} increase the LLM's ability to understand the code. Again a bootstrap confirms the proportion of errors and failed tests did not change between the \textit{base file} with/without comments by a statistically significant margin.

\begin{table*}[t]
    \centering
    \begin{tabular}{c|c|c|c|c|c}
         Model&Basefile&Comments&Random Comments&Random Names&Animal Names \\ \hline
         GPT-3.5&1323 &883 &892 &1464 &1482  \\ \hline
         GPT-4&3832 &3295 &2324 &3725 &3589 \\
    \end{tabular}
    \caption{Number of tests generated by models under variations of HumanEval files.}
    \label{tab:cat_num_tests}
\end{table*}

\begin{table*}[t]
    \centering
    \begin{tabular}{c|c|c|c|c|c|c}
         Model&p = 0&p = 0.1& p = 0.25 & p = 0.50&p = 0.75&p = 1 \\ \hline
         GPT-3.5&1323 &958 &929 &893 &908&883  \\ \hline
         GPT-4&3832 &3596 &3446 &3198 &3335&3295 \\
    \end{tabular}
    \caption{Number of tests generated by models with proportions of dropped comment lines within HumanEval files.}
    \label{tab:drop_num_tests}
\end{table*}

\begin{figure}
    \centering
    \includegraphics[width=\linewidth]{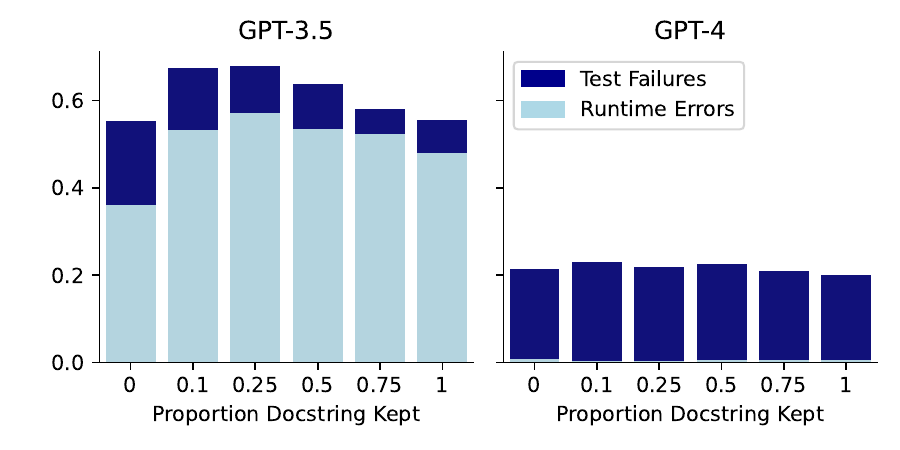}
    \caption{Proportion of runtime errors or failed tests that happen with GPT-3.5 (left) and GPT-4 (right) generating unit tests on different proportions of docstring lines kept on HumanEval code.}
    \label{fig:humaneval_drop}
\end{figure}

As mentioned above, we also conducted experiments to determine how partial comments might affect an LLM's ability to understand code. Figure~\ref{fig:humaneval_drop} shows these results. GPT-3.5 shows that providing partial comments causes an initial spike in the number of errors, before declining as the percentage of comments kept increases. GPT-4 does not show the same pattern, however a bootstrap shows that having the full amount of comments gave fewer errors over when the proportion of kept comments was 10\%, 25\% and 50\% by statistically significant margin. As shown in the figure, these changes were relatively minor. No other comparisons were statistically significant for GPT-4. We therefore cannot currently conclude whether or not partial comments can significantly hurt an LLM's code understanding based on the current evidence.

\begin{figure}
    \centering
    \includegraphics[width=\linewidth]{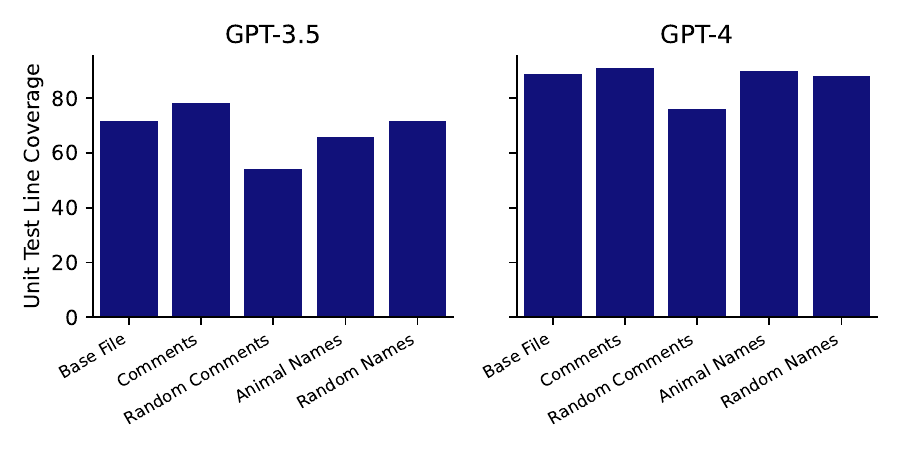}
    \caption{Average percent of line coverage with GPT-3.5 (left) and GPT-4 (right) generating unit tests on modified versions of HumanEval code.}
    \label{fig:base_line_coverage}
\end{figure}

\begin{figure}
    \centering
    \includegraphics[width=\linewidth]{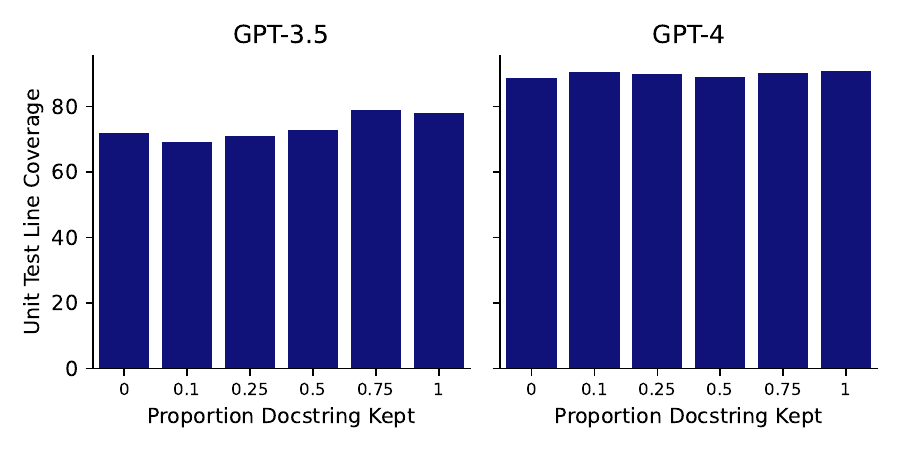}
    \caption{Average percent of line coverage with GPT-3.5 (left) and GPT-4 (right) generating unit tests on different proportions of docstring lines kept on HumanEval code.}
    \label{fig:coment_drop_line_coverage}
\end{figure}

Finally, we conducted an analysis of line coverage by unit tests to prevent the possibility of the LLM producing trivial tests to gain a higher success rate. This is shown in Figures~\ref{fig:base_line_coverage}  and \ref{fig:coment_drop_line_coverage}. First, it is worth noting that the results with \textit{random comments} had lower average code coverage than all other methods by a statistically significant margin. Second, we note that for both GPT-3.5 and GPT-4, code with \textit{comments} generated a higher average line coverage than all other methods by a statistically significant margin. Thirdly, we note that modifying the variables names of the code largely did not have a statistically significant effect. The one exception was the \textit{animal names} category generated by GPT-3.5, which had an average amount of line coverage that was significantly lower than the line coverage of unit tests generated from the baseline, as seen in Figure~\ref{fig:base_line_coverage}. Finally, partial docstrings mostly did not have a statistically significant effect on the average amount of line coverage. There was one exception when 10\% of the docstring was kept for GPT-3.5, which did worse than the baseline by a statistically significant margin.

We also provide the number of unit tests generated by models for various conditions. Table~\ref{tab:cat_num_tests} shows the number of test cases created by GPT-3.5 and GPT-4 for each of the file categories mentioned above. Table~\ref{tab:drop_num_tests} shows the number of tests generated by GPT-3.5 and GPT-4 with various portions of the HumanEval docstring lines dropped.

%While there are differences in results across different comment proportions for GPT-3.5, there does not seem to be a clear correlation when comparing the amount of comments to the proportion of successfully run unit tests. There is no statistically significant differences for GPT-4 with 2 exceptions.  We therefore conclude that correct but incomplete comments do not significantly impact an LLM's code understanding.
% TODO: confirm significance testing.

\section{Conclusions and Discussion}

% Conclude
In this paper, we introduce the effect of code documentation on LLM code understanding and show in our initial experiments that the relative prevalence of documentation has little to no significant effect on an LLM's understanding as we have defined it. This is a little complicated by the fact that we found no significant difference in the amount of successful unit tests but did see a significant difference in the code coverage of those unit tests. Even when comparing commented code to code without comments and all of the variable names changed to random characters, we find little to no significant difference in the LLM's ability to generate successful unit tests. So, although more of the code with \textit{comments} is being covered by the unit tests than the code without, this does not improve the overall unit test success rate. This suggests that the LLM is better at understanding the different execution paths of the code, but this may make the creation of a successful unit test more difficult. Alternatively, we do show that incorrect comments do significantly affect an LLM's ability to understand a piece of code.

% Discuss
It is possible that in much of the training data used to train the OpenAI models tested in this study, the code did not contain many comments or documentation. As such, it may be possible that a model's ability to utilize comments is dependent on how much of this documentation is in the training data, and other non-OpenAI LLM's may make better or worse use of the information provided. It is also possible that correct comments do not add much information as it would be relevant to the creation of unit tests, but incorrect comments still \textit{confuse} the LLMs.

% \md{\begin{enumerate}
%     \item why we think variable names did not change ability
%     \item why we think code comments did not improve output significantly (was it significant from random variable names?)
% \end{enumerate}}

\section{Limitations}
%Limitations and future work
While we do introduce a new research question to the space of LLM code understanding, we recognize that this is a limited test that is missing a broader comparison study between programming languages, models, prompting techniques, hyperparameters, and input documentation modification, and that this paper represents merely an introduction to the idea of documentation's effects on LLM code understanding. In addition, since OpenAI's training data is not publicly available, there is a small but real possibility that the training data may contain the HumanEval dataset, in which case these results would be invalid.

In future work, we believe that more comprehensive tests should be done by including examples of more complex code and documentation (such as that found in ClassEval~\cite{du2023classeval}). These tests could be evaluated while simultaneously seeing the affect of and  controlling for code complexity using metrics such as cyclomatic complexity or maintainability index~\cite{1702388, 303623}. We also recommend evaluating unit test success rate while controlling for code coverage to properly isolate and determine good success metrics.

Finally, this same research question could be evaluated using code summarization as the code understanding task, because it still measures how well the LLM attempts to answer ``What does this piece of code do?''

% UNCOMMENT before publishing
\section*{Acknowledgments}
The authors thank Tim Welsh and Guido Zarrella for advice on experimental design and editing. This work was funded under MITRE’s 2023 Independent Research and Development Program.

\bibliography{custom}

\begin{thebibliography}{33}
\expandafter\ifx\csname natexlab\endcsname\relax\def\natexlab#1{#1}\fi

\bibitem[{Agnihotri and Chug()}]{agnihotri2020systematic}
Mansi Agnihotri and Anuradha Chug.
\newblock A systematic literature survey of software metrics, code smells and refactoring techniques.
\newblock \emph{Journal of Information Processing Systems}, 16(4):915--934.

\bibitem[{Bui et~al.(2023)Bui, Le, Wang, Li, Gotmare, and Hoi}]{bui2023codetf}
Nghi D.~Q. Bui, Hung Le, Yue Wang, Junnan Li, Akhilesh~Deepak Gotmare, and Steven C.~H. Hoi. 2023.
\newblock \href {http://arxiv.org/abs/2306.00029} {Codetf: One-stop transformer library for state-of-the-art code llm}.

\bibitem[{Chen et~al.(2021)Chen, Tworek, Jun, Yuan, de~Oliveira~Pinto, Kaplan, Edwards, Burda, Joseph, Brockman, Ray, Puri, Krueger, Petrov, Khlaaf, Sastry, Mishkin, Chan, Gray, Ryder, Pavlov, Power, Kaiser, Bavarian, Winter, Tillet, Such, Cummings, Plappert, Chantzis, Barnes, Herbert-Voss, Guss, Nichol, Paino, Tezak, Tang, Babuschkin, Balaji, Jain, Saunders, Hesse, Carr, Leike, Achiam, Misra, Morikawa, Radford, Knight, Brundage, Murati, Mayer, Welinder, McGrew, Amodei, McCandlish, Sutskever, and Zaremba}]{chen2021codex}
Mark Chen, Jerry Tworek, Heewoo Jun, Qiming Yuan, Henrique~Ponde de~Oliveira~Pinto, Jared Kaplan, Harri Edwards, Yuri Burda, Nicholas Joseph, Greg Brockman, Alex Ray, Raul Puri, Gretchen Krueger, Michael Petrov, Heidy Khlaaf, Girish Sastry, Pamela Mishkin, Brooke Chan, Scott Gray, Nick Ryder, Mikhail Pavlov, Alethea Power, Lukasz Kaiser, Mohammad Bavarian, Clemens Winter, Philippe Tillet, Felipe~Petroski Such, Dave Cummings, Matthias Plappert, Fotios Chantzis, Elizabeth Barnes, Ariel Herbert-Voss, William~Hebgen Guss, Alex Nichol, Alex Paino, Nikolas Tezak, Jie Tang, Igor Babuschkin, Suchir Balaji, Shantanu Jain, William Saunders, Christopher Hesse, Andrew~N. Carr, Jan Leike, Josh Achiam, Vedant Misra, Evan Morikawa, Alec Radford, Matthew Knight, Miles Brundage, Mira Murati, Katie Mayer, Peter Welinder, Bob McGrew, Dario Amodei, Sam McCandlish, Ilya Sutskever, and Wojciech Zaremba. 2021.
\newblock \href {http://arxiv.org/abs/2107.03374} {Evaluating large language models trained on code}.

\bibitem[{Coleman et~al.(1994)Coleman, Ash, Lowther, and Oman}]{303623}
D.~Coleman, D.~Ash, B.~Lowther, and P.~Oman. 1994.
\newblock \href {https://doi.org/10.1109/2.303623} {Using metrics to evaluate software system maintainability}.
\newblock \emph{Computer}, 27(8):44--49.

\bibitem[{Du et~al.(2023)Du, Liu, Wang, Wang, Liu, Chen, Feng, Sha, Peng, and Lou}]{du2023classeval}
Xueying Du, Mingwei Liu, Kaixin Wang, Hanlin Wang, Junwei Liu, Yixuan Chen, Jiayi Feng, Chaofeng Sha, Xin Peng, and Yiling Lou. 2023.
\newblock \href {http://arxiv.org/abs/2308.01861} {Classeval: A manually-crafted benchmark for evaluating llms on class-level code generation}.

\bibitem[{Hong et~al.(2023)Hong, Zhuge, Chen, Zheng, Cheng, Zhang, Wang, Wang, Yau, Lin, Zhou, Ran, Xiao, Wu, and Schmidhuber}]{hong2023metagpt}
Sirui Hong, Mingchen Zhuge, Jonathan Chen, Xiawu Zheng, Yuheng Cheng, Ceyao Zhang, Jinlin Wang, Zili Wang, Steven Ka~Shing Yau, Zijuan Lin, Liyang Zhou, Chenyu Ran, Lingfeng Xiao, Chenglin Wu, and Jürgen Schmidhuber. 2023.
\newblock \href {http://arxiv.org/abs/2308.00352} {Metagpt: Meta programming for a multi-agent collaborative framework}.

\bibitem[{Jiao et~al.(2023)Jiao, Wang, Huang, Wang, Shi, and Tu}]{Jiao2023-xp}
Wenxiang Jiao, Wenxuan Wang, Jen-Tse Huang, Xing Wang, Shuming Shi, and Zhaopeng Tu. 2023.
\newblock \href {http://arxiv.org/abs/2301.08745} {Is {ChatGPT} a good translator? yes with {GPT-4} as the engine}.

\bibitem[{Kalliamvakou(2022)}]{githubResearchQuantifying}
Eirini Kalliamvakou. 2022.
\newblock {R}esearch: quantifying {G}it{H}ub {C}opilot’s impact on developer productivity and happiness --- github.blog.
\newblock \url{https://github.blog/2022-09-07-research-quantifying-github-copilots-impact-on-developer-productivity-and-happiness/}.
\newblock [Accessed 27-11-2023].

\bibitem[{Kocmi and Federmann(2023)}]{kocmi-federmann-2023-large}
Tom Kocmi and Christian Federmann. 2023.
\newblock \href {https://aclanthology.org/2023.eamt-1.19} {Large language models are state-of-the-art evaluators of translation quality}.
\newblock In \emph{Proceedings of the 24th Annual Conference of the European Association for Machine Translation}, pages 193--203, Tampere, Finland. European Association for Machine Translation.

\bibitem[{Le et~al.(2023)Le, Chen, Saha, Gokul, Sahoo, and Joty}]{le2023codechain}
Hung Le, Hailin Chen, Amrita Saha, Akash Gokul, Doyen Sahoo, and Shafiq Joty. 2023.
\newblock \href {http://arxiv.org/abs/2310.08992} {Codechain: Towards modular code generation through chain of self-revisions with representative sub-modules}.

\bibitem[{Leinonen et~al.(2023)Leinonen, Denny, MacNeil, Sarsa, Bernstein, Kim, Tran, and Hellas}]{leinonen2023comparing}
Juho Leinonen, Paul Denny, Stephen MacNeil, Sami Sarsa, Seth Bernstein, Joanne Kim, Andrew Tran, and Arto Hellas. 2023.
\newblock Comparing code explanations created by students and large language models.
\newblock \emph{arXiv preprint arXiv:2304.03938}.

\bibitem[{Lenarduzzi et~al.(2020)Lenarduzzi, Sillitti, and Taibi}]{lenarduzzi2020survey}
Valentina Lenarduzzi, Alberto Sillitti, and Davide Taibi. 2020.
\newblock A survey on code analysis tools for software maintenance prediction.
\newblock In \emph{Proceedings of 6th International Conference in Software Engineering for Defence Applications: SEDA 2018 6}, pages 165--175. Springer.

\bibitem[{Lu et~al.(2021)Lu, Guo, Ren, Huang, Svyatkovskiy, Blanco, Clement, Drain, Jiang, Tang, Li, Zhou, Shou, Zhou, Tufano, Gong, Zhou, Duan, Sundaresan, Deng, Fu, and Liu}]{lu2021codexglue}
Shuai Lu, Daya Guo, Shuo Ren, Junjie Huang, Alexey Svyatkovskiy, Ambrosio Blanco, Colin Clement, Dawn Drain, Daxin Jiang, Duyu Tang, Ge~Li, Lidong Zhou, Linjun Shou, Long Zhou, Michele Tufano, Ming Gong, Ming Zhou, Nan Duan, Neel Sundaresan, Shao~Kun Deng, Shengyu Fu, and Shujie Liu. 2021.
\newblock \href {http://arxiv.org/abs/2102.04664} {Codexglue: A machine learning benchmark dataset for code understanding and generation}.

\bibitem[{Ma et~al.(2023)Ma, Liu, Wang, Hu, Liu, Zhang, Nie, and Liu}]{ma2023chatgpt}
Wei Ma, Shangqing Liu, Wenhan Wang, Qiang Hu, Ye~Liu, Cen Zhang, Liming Nie, and Yang Liu. 2023.
\newblock \href {http://arxiv.org/abs/2305.12138} {Chatgpt: Understanding code syntax and semantics}.

\bibitem[{McCabe(1976)}]{1702388}
T.J. McCabe. 1976.
\newblock \href {https://doi.org/10.1109/TSE.1976.233837} {A complexity measure}.
\newblock \emph{IEEE Transactions on Software Engineering}, SE-2(4):308--320.

\bibitem[{Moor et~al.(2007)Moor, Verbaere, Hajiyev, Avgustinov, Ekman, Ongkingco, Sereni, and Tibble}]{4362893}
Oege~de Moor, Mathieu Verbaere, Elnar Hajiyev, Pavel Avgustinov, Torbjorn Ekman, Neil Ongkingco, Damien Sereni, and Julian Tibble. 2007.
\newblock \href {https://doi.org/10.1109/SCAM.2007.31} {Keynote address: .ql for source code analysis}.
\newblock In \emph{Seventh IEEE International Working Conference on Source Code Analysis and Manipulation (SCAM 2007)}, pages 3--16.

\bibitem[{Moslem et~al.(2023)Moslem, Haque, Kelleher, and Way}]{moslem-etal-2023-adaptive}
Yasmin Moslem, Rejwanul Haque, John~D. Kelleher, and Andy Way. 2023.
\newblock \href {https://aclanthology.org/2023.eamt-1.22} {Adaptive machine translation with large language models}.
\newblock In \emph{Proceedings of the 24th Annual Conference of the European Association for Machine Translation}, pages 227--237, Tampere, Finland. European Association for Machine Translation.

\bibitem[{{OpenAI}(2023)}]{OpenAI2023-xj}
{OpenAI}. 2023.
\newblock \href {http://arxiv.org/abs/2303.08774} {{GPT-4} technical report}.

\bibitem[{Palacio et~al.(2023)Palacio, Velasco, Rodriguez-Cardenas, Moran, and Poshyvanyk}]{palacio2023evaluating}
David~N Palacio, Alejandro Velasco, Daniel Rodriguez-Cardenas, Kevin Moran, and Denys Poshyvanyk. 2023.
\newblock \href {http://arxiv.org/abs/2308.03873} {Evaluating and explaining large language models for code using syntactic structures}.

\bibitem[{Pan et~al.(2023)Pan, Ibrahimzada, Krishna, Sankar, Wassi, Merler, Sobolev, Pavuluri, Sinha, and Jabbarvand}]{Pan2023-rx}
Rangeet Pan, Ali~Reza Ibrahimzada, Rahul Krishna, Divya Sankar, Lambert~Pouguem Wassi, Michele Merler, Boris Sobolev, Raju Pavuluri, Saurabh Sinha, and Reyhaneh Jabbarvand. 2023.
\newblock \href {http://arxiv.org/abs/2308.03109} {Understanding the effectiveness of large language models in code translation}.

\bibitem[{Pritchard(2014)}]{Pritchard2014}
Duncan Pritchard. 2014.
\newblock \href {https://doi.org/10.1007/978-3-319-04672-3_18} {\emph{Knowledge and Understanding}}, page 315–327. Springer International Publishing.

\bibitem[{Pritchard(2016)}]{Pritchard2016-PRISIF}
Duncan Pritchard. 2016.
\newblock \href {https://doi.org/10.1017/epi.2015.59} {Seeing it for oneself: Perceptual knowledge, understanding, and intellectual autonomy}.
\newblock \emph{Episteme}, 13(1):29--42.

\bibitem[{Sch{\"a}fer et~al.(2023)Sch{\"a}fer, Nadi, Eghbali, and Tip}]{schafer2023empirical}
Max Sch{\"a}fer, Sarah Nadi, Aryaz Eghbali, and Frank Tip. 2023.
\newblock An empirical evaluation of using large language models for automated unit test generation.
\newblock \emph{arXiv preprint arXiv:2302.06527}.

\bibitem[{Shinn et~al.(2023)Shinn, Cassano, Berman, Gopinath, Narasimhan, and Yao}]{shinn2023reflexion}
Noah Shinn, Federico Cassano, Edward Berman, Ashwin Gopinath, Karthik Narasimhan, and Shunyu Yao. 2023.
\newblock \href {http://arxiv.org/abs/2303.11366} {Reflexion: Language agents with verbal reinforcement learning}.

\bibitem[{Solar-Lezama(2009)}]{solar2009sketching}
Armando Solar-Lezama. 2009.
\newblock The sketching approach to program synthesis.
\newblock In \emph{Asian symposium on programming languages and systems}, pages 4--13. Springer.

\bibitem[{Sun et~al.(2023)Sun, Fang, You, Miao, Liu, Li, Deng, Huang, Chen, Zhang, Qian, Liu, and Chen}]{Sun2023-sp}
Weisong Sun, Chunrong Fang, Yudu You, Yun Miao, Yi~Liu, Yuekang Li, Gelei Deng, Shenghan Huang, Yuchen Chen, Quanjun Zhang, Hanwei Qian, Yang Liu, and Zhenyu Chen. 2023.
\newblock \href {http://arxiv.org/abs/2305.12865} {Automatic code summarization via {ChatGPT}: How far are we?}

\bibitem[{Tim{\'o}teo et~al.(2008)Tim{\'o}teo, {\'A}lvaro, De~Almeida, and de~Lemos~Meira}]{timoteo2008software}
Aline~Lopes Tim{\'o}teo, Alexandre {\'A}lvaro, Eduardo~Santana De~Almeida, and Silvio~Romero de~Lemos~Meira. 2008.
\newblock Software metrics: A survey.
\newblock \emph{Sl: sn}.

\bibitem[{Wang et~al.(2022)Wang, Yang, Gao, Peng, Zhang, and Lyu}]{10.1145/3540250.3549113}
Chaozheng Wang, Yuanhang Yang, Cuiyun Gao, Yun Peng, Hongyu Zhang, and Michael~R. Lyu. 2022.
\newblock \href {https://doi.org/10.1145/3540250.3549113} {No more fine-tuning? an experimental evaluation of prompt tuning in code intelligence}.
\newblock In \emph{Proceedings of the 30th ACM Joint European Software Engineering Conference and Symposium on the Foundations of Software Engineering}, ESEC/FSE 2022, page 382–394, New York, NY, USA. Association for Computing Machinery.

\bibitem[{Wang et~al.(2023{\natexlab{a}})Wang, Liu, Wang, Cui, Ding, Liu, and Yu}]{wang2023intervenor}
Hanbin Wang, Zhenghao Liu, Shuo Wang, Ganqu Cui, Ning Ding, Zhiyuan Liu, and Ge~Yu. 2023{\natexlab{a}}.
\newblock Intervenor: Prompt the coding ability of large language models with the interactive chain of repairing.
\newblock \emph{arXiv preprint arXiv:2311.09868}.

\bibitem[{Wang et~al.(2023{\natexlab{b}})Wang, Le, Gotmare, Bui, Li, and Hoi}]{wang2023codet5+}
Yue Wang, Hung Le, Akhilesh~Deepak Gotmare, Nghi~DQ Bui, Junnan Li, and Steven~CH Hoi. 2023{\natexlab{b}}.
\newblock Codet5+: Open code large language models for code understanding and generation.
\newblock \emph{arXiv preprint arXiv:2305.07922}.

\bibitem[{Xia et~al.(2022)Xia, Wei, and Zhang}]{xia2022practical}
Chunqiu~Steven Xia, Yuxiang Wei, and Lingming Zhang. 2022.
\newblock \href {http://arxiv.org/abs/2210.14179} {Practical program repair in the era of large pre-trained language models}.

\bibitem[{Xu et~al.(2022)Xu, Alon, Neubig, and Hellendoorn}]{xu2022systematic}
Frank~F Xu, Uri Alon, Graham Neubig, and Vincent~Josua Hellendoorn. 2022.
\newblock A systematic evaluation of large language models of code.
\newblock In \emph{Proceedings of the 6th ACM SIGPLAN International Symposium on Machine Programming}, pages 1--10.

\bibitem[{Zelikman et~al.(2023)Zelikman, Huang, Poesia, Goodman, and Haber}]{zelikman2023parsel}
Eric Zelikman, Qian Huang, Gabriel Poesia, Noah Goodman, and Nick Haber. 2023.
\newblock Parsel: Algorithmic reasoning with language models by composing decompositions.
\newblock In \emph{Thirty-seventh Conference on Neural Information Processing Systems}.

\end{thebibliography}

\appendix

\end{document}